%% file: main.tex
\documentclass[submission,copyright,creativecommons]{eptcs}
\providecommand{\event}{ALF 2014} 
\usepackage{breakurl}             
\usepackage{algorithmicx}
\usepackage{algpseudocode}

\input{macros-common.tex}

\input{macros.tex}

\newtheorem{theorem}{Theorem}

\newtheorem{remark}{Remark}

\title{Saturation algorithms for model-checking pushdown systems\thanks{We thank
Didier Caucal and Olivier Serre for helpful discussions.  This survey was supported by the
Engineering and Physical Sciences Research Council [EP/K009907/1].}}

\author{Arnaud Carayol
\institute{LIGM}
\institute{Universit\'e Paris-Est \& CNRS}
\email{arnaud.carayol@univ-mlv.fr}
\and
Matthew Hague
\institute{Department of Computer Science}
\institute{ Royal Holloway University of London}
\email{matthew.hague@rhul.ac.uk}
}
\def\titlerunning{Saturation algorithms for model-checking pushdown systems}
\def\authorrunning{A. Carayol and M. Hague}
\begin{document}
\maketitle

\begin{abstract}
We present a survey of the saturation method for model-checking pushdown systems.
\end{abstract}

\input{introduction.tex}

\input{preliminaries.tex}

\input{reachability.tex}

\input{games.tex}

\input{implementations.tex}

\input{others.tex}

\bibliographystyle{eptcs}
\bibliography{afl}

\end{document}

%% file: macros-common.tex
\newcommand\astates{\mathbb{S}}
\newcommand\afinals{\mathcal{F}}
\newcommand\ainits{\mathcal{I}}
\newcommand\adelta{\delta}

%% file: macros.tex
\usepackage{amsfonts}
\usepackage{amsmath}
\usepackage{amssymb}
\usepackage{xspace}
\usepackage{stmaryrd}
\usepackage{pifont}

\newcommand\neworrenewcommand[1]{%
    \let#1\relax
    \newcommand#1%
}

\newcommand\cshore{C-SHORe\xspace}

\newcommand\horsat{HorSat\xspace}

\newcommand\etal{~\textit{et al.}\xspace}

\newcommand\setcomp[2]{\left\{{#1}\ \left|\ {#2}\right.\right\}}
\newcommand\set[1]{\left\{{#1}\right\}}

\newcommand\brac[1]{\left({#1}\right)}
\newcommand\apply[2]{{#1}\mathord{\brac{#2}}}

\neworrenewcommand\ordinal{\alpha}

\newcommand\negation[1]{\overline{#1}}

\newcommand\buchi{B\"uchi\xspace}

\newcommand\prestar[2]{\apply{Pre^*_{#1}}{#2}}
\newcommand\poststar[2]{\apply{Post^*_{#1}}{#2}}

\newcommand\alphabet\Sigma

\newcommand\oalphabet\Gamma
\newcommand\ocha\gamma

\newcommand\lang{\mathcal{L}}
\newcommand\langof[1]{\apply{\lang}{#1}}
\newcommand\slang[2]{\apply{\lang_{#1}}{#2}}

\newcommand{\abelard}{Abelard\xspace}
\newcommand{\eloise}{\'Eloise\xspace}

\newcommand\winningregion{\mathcal{W}}
\newcommand\wincond{W}



%% file: introduction.tex
\section{Introduction}

Pushdown systems have, over the past 15 years, been popular with the
software verification community.  Their stack can be used to model the call
stack of a first-order recursive program, with the control state holding
valuations of the program's global variables, and stack characters encoding the
local variable valuations.  As such the control flow of first-order recursive
programs (such as C and Java programs) can be accurately modelled~\cite{JM77}.  Pushdown systems have played a key role in the automata-theoretic approach to software
model checking and considerable progress has been made in the implementation of
scalable model checkers of pushdown systems.  These tools (e.g.
Bebop~\cite{BR00} and Moped~\cite{ES01,S02,SSE05,SBSE07}) are an essential
back-end components of high-profile model checkers such as SLAM~\cite{BLR11}.

A fundamental result for the model-checking of pushdown systems  was established by \buchi in \cite{Buchi64}. He showed that the set of stack contents reachable from the initial configuration of a pushdown system form a regular language and hence can be represented by a finite state automaton. The procedure provided by \buchi to compute this automaton from the pushdown system is exponential. In \cite{Caucal88}, Caucal gave the first polynomial time algorithm to solve this problem. This efficient computation is obtained by a saturation process where transitions are incrementally added to the finite automaton. This technique, which is the topic of this survey, was simplified and adapted to the model-checking setting by Bouajjani\etal in \cite{BEM97} and independently by Finkel\etal in \cite{FWW97}.

The saturation technique allows global model checking of pushdown systems.  For
example, one may construct a regular representation of all configurations
reachable from a given set of initial configurations, or, dually, the set of
all configurations that may reach a given set of target configurations.  As
well as providing direct solutions to simple reachability properties (e.g.  can
an error state be reached from a designated initial configuration), the
representations constructed by global analyses may be reused in a variety of
settings.  For example, once may perform multiple (and dynamic) queries on the
set of reachable states without having to re-run the model checking routine.
Additionally, these representations may be combined as part of a larger
algorithm or proof.  For example, Bouajjani\etal provided solutions to the
model checking problem for the alternation free $\mu$-calculus by
combining the results obtained through multiple global reachability
analyses~\cite{BEM97}.

In this survey, we present the saturation method under its different forms
for reachability problems in Section~\ref{sec:reachability}. The saturation technique also generalises to the analysis of two-players games played over the configuration graph of a pushdown systems. This extension based on the work of Cachat \cite{Cachat02} and Hague and Ong \cite{HagueO09} is presented in Section~\ref{sec:games}. In Section~\ref{sec:implementations}, we review the various model-checking tools that implement the saturation technique. We conclude in Section~\ref{sec:others} by giving an overview of the extensions of the basic model of pushdown system for which the saturation technique has been applied.


%% file: preliminaries.tex
\section{Preliminaries}
\label{sec:preliminaries}

\newcommand{\pda}{P}
\newcommand{\paut}{\mathcal{A}}
\newcommand{\pautb}{\mathcal{B}}
\newcommand{\era}[1]{\xrightarrow{#1}}
\newcommand{\eRb}[2]{\overset{#1}{\underset{#2}{\Longrightarrow}}}
\newcommand{\erb}[2]{\xrightarrow[#2]{#1}}
\newcommand{\eVb}[2]{\overset{#1}{\underset{#2}{\leadsto}}}

\subsection{Finite automata}

We denote by $\Sigma^{*}$ the set of words over the finite alphabet $\Sigma$.
For $n \geq 0$, we denote by $\Gamma^{\leq n}$ the set of words of length at most $n$.

A finite automaton $\mathcal{A}$ over the alphabet $\Sigma$ is a tuple $(\astates,\ainits,\afinals,\adelta)$
where $\astates$ is a finite set of states, $\ainits \subseteq \astates$ is the set of initial states, $\afinals \subseteq \astates$ is the set of final states and $\adelta \subseteq \astates \times \Sigma \times \astates$ is the set of transitions. We write $s \erb{a}{\mathcal{A}} t$ to denote that $(s,a,t)$ is a transition of $\mathcal{A}$. For a word $w \in \Sigma^{*}$, we write $s \eRb{w}{\mathcal{A}} t$ to denote the fact that $\mathcal{A}$ can reach the state $t$ while reading the word $w$ starting from the state $s$.
The language accepted by $\mathcal{A}$ from a state $s$ is
\[
    \slang{s}{\mathcal{A}} = \setcomp{w \in \Sigma^\ast}{\exists s_{f} \in \afinals . s \eRb{w}{\mathcal{A}} s_f}
\]
and the language accepted by $\mathcal{A}$ is
\[
    \langof{\mathcal{A}} = \bigcup\limits_{s \in \ainits} \slang{s}{\mathcal{A}} \ .
\]

\subsection{Pushdown system}

A pushdown system $\pda$ is a given by a tuple $(Q,\Gamma,\bot,\Delta)$ where  $Q$ is a finite set of control states, $\Gamma$ is the finite stack alphabet, $\bot \in \Gamma$ is a special bottom of stack symbol  and $\Delta \subseteq (Q \times \Gamma) \times (Q \times \Gamma^{\leq 2})$ is the set of transitions. We write $(q,A) \rightarrow (p,w)$ for the transition $((q,A),(p,w))$. A configuration is a tuple $(q,w)$ where $q$ is a state in $Q$ and $w$ is a stack content in $(\Gamma \setminus \{\bot\})^{*}\bot$. In the configuration $c=(q,Aw)$, the pushdown system can apply the transition $(q,A) \rightarrow (p,u)$ to go to the configuration $c'=
(p,uw)$. As is usual, we assume that transitions of the pushdown system does not pop the bottom of stack symbol or does not push it on the stack (\emph{i.e.} all transitions involving the symbol $\bot$ are of the form
$q\bot \rightarrow p\bot$ or $q\bot \rightarrow p\bot A$ for some $A \in \Gamma \setminus \{\bot\}$). We denote by $\erb{}{\pda}$ (or simply $\rightarrow$ if $\pda$ is clear from the context) the relation on configurations defined by the transitions of $\pda$. We denote by $\eRb{}{\pda}$ the reflexive and transitive closure of $\erb{}{\pda}$.

%% file: reachability.tex
\section{Reachability problems for pushdown systems}
\label{sec:reachability}

A fundamental result for the model-checking of pushdown systems is the fact
that the set of stack contents:
\[
\{ w \in \Gamma^{*} \mid  \exists q \in Q, (q_{0},\bot) \Rightarrow (q,w) \}
\]
\noindent
that are reachable from an arbitrary initial configuration of the system, form a regular set of words over the stack alphabet $\Gamma$.

A more elegant formulation of this result can be obtained by extending the notion of regularity to sets of configurations. A set of configurations $C$ is \emph{regular} if for every state $p \in Q$, the set of associated stack contents $\{ w \in \Gamma^{*} \mid (p,w) \in C\}$ is regular. A $\pda$-automaton
is a slight extension of the standard notion of finite automaton to accept
configurations. The only extra assumption is that the set of states of the $\pda$-automaton contains the set of states of the pushdown system. Formally, a $\pda$-automaton is of the form $(\astates,Q,\afinals,\adelta)$ where $Q$ is the set
of states of the pushdown system $\pda$. A configuration is $(p,w)$ is accepted by $\paut$ if $w$ is accepted by $\paut$ starting from the state $p$ (\emph{i.e.} $w \in L_{p}(\paut)$).

\begin{theorem}{\cite{Buchi64}}
\label{thm:buchi}
The set of configurations of a pushdown system reachable from the initial configuration (\emph{i.e.} the configuration $(q_{0},\bot)$ for some arbitrary state $q_{0}$) is regular. Moreover a $\pda$-automaton accepting it can be effectively constructed from the pushdown system.
\end{theorem}

To the authors knowledge, the first  proof of this result is due to \buchi in \cite{Buchi64}. The formalism used by \buchi is not that of pushdown automata
but that of prefix word-rewriting systems (which he calls \emph{regular canonical systems}). These systems syntactically include pushdown automata and conversely can be simulated by pushdown automata. In \cite{Greibach67}, Greibach formalises the correspondence between the two models and gives a simple proof based on a result on context-free languages proved by Bar-Hillel \emph{et al.} in \cite{Bar-hillel61}. Greibach also says that the result (for pushdown automata) was part of the folklore at the time but never appeared in print. Even though effective, these proofs do not provide a polynomial time algorithm\footnote{We will see Section~\ref{ssec:buchi-proof} that it can easily be adapted to provide a polynomial time algorithm.}. The first polynomial time algorithm is due to Caucal \cite{Caucal88,Caucal90} which is based on a saturation procedure of a finite state automaton. The idea behind the saturation method can be traced back to \cite{Benois69}. This method was independently rediscovered and used for model-checking purposes by Bouajjani \emph{et al.} in \cite{BEM97} and Finkel \emph{et al.} in \cite{FWW97}.

A more general problem is, given a regular set of configurations $C$, to compute the set:
\[
\poststar{\pda}{C} = \{ c'  \mid \exists c \in C, c \eRb{}{\pda} c' \}
\]
\noindent
of configurations that can be reached from a configuration in $C$.

The regularity of $\poststar{}{C}$, for any regular set $C$, can  be derived from Theorem~\ref{thm:buchi}. Indeed starting from a pushdown system $\pda$ and a regular set of configurations $C$, we can create a new pushdown system $\pda'$ which using new states builds any configuration in $C$ and afterwards behaves like $\pda$. Clearly the set of configurations reachable from the initial configuration of $\pda'$ coincide with $\poststar{\pda}{C}$ when restricted to
the states of $\pda$.

As mentioned in the introduction, for model-checking purposes it is often interesting to compute the set of configurations that can reach a given set of \emph{bad} configurations. This leads to consider the set
\[
\prestar{\pda}{C} = \{ c'  \mid \exists c \in C, c' \Rightarrow c \}
\]
\noindent
of configurations that can reach a configuration in $C$.

The regularity of $\prestar{}{C}$ for any regular set $C$ can be deduced from the regularity of $\poststar{}{C}$.
The intuitive idea is to construct, from $\pda$, a new pushdown system $\pda'$ whose derivation relation is the inverse
of that of $\pda$. For a transition of the form $qA \rightarrow p$ of $\pda$, we add the transitions $p X \rightarrow q AX$
for all symbols $X \in \Gamma$. For a transition $qA \rightarrow pBC$ of $\pda$, we add two transition $pB \rightarrow r_{(C,q,A)}$
and $r_{(C,q,A)} C  \rightarrow  q A$ where $r_{(C,q,A)}$ is a new intermediary control state. For any two configurations $c$ and $c'$ of $\pda$,
it holds that $c \Rightarrow_{\pda} c'$ if and only if $c' \Rightarrow_{\pda'} c$. Hence $\prestar{\pda}{C}$ is equal to the restriction
of $\poststar{\pda'}{C}$ to the states of $\pda$ and is therefore regular.

The section is structured as follows. We present \buchi's original proof in Section~\ref{ssec:buchi-proof}. In Section~\ref{ssec:saturation}, we present the saturation algorithm to compute $\prestar{}{C}$ introduced in \cite{BEM97}. Finally in Section~\ref{ssec:derivation}, we characterise the derivation relation of the pushdown automata using the saturation technique following \cite{Caucal88}.

\subsection{\buchi's proof}
\label{ssec:buchi-proof}

We present a proof of Theorem~\ref{thm:buchi} adapted from \cite{Buchi64}. In the original proof, \buchi first reduced the problem to a very simple form of pushdown system where transitions are either of the form $pA \rightarrow q$ or $p \rightarrow qA$. This model (called \emph{reduced regular systems} by \buchi) is completely symmetric and therefore computing $Pre^{*}$ or $Post^{*}$ is essentially the same thing. However to adapt the proof to the formalism used in this article (recall that our formalism does not allow rules of the form $p \rightarrow qA$), it is more convenient to work with $Pre^{*}$ than with $Post^{*}$.

Given a pushdown system $\pda=(Q,F,\bot,\Delta)$, we construct a $\pda$-automaton accepting $\prestar{\pda}{\{(q_{f},\bot)\}}$ where $q_{f}$ is an arbitrary \emph{final} state of the pushdown system.

The construction is based on the following remark: to reach
the configuration $(q_{f},\bot)$ from a configuration $(p,Aw\bot)$
it is necessary, at some point, to reach a configuration of the form $(q,w\bot)$ for
some state $q \in Q$. Moreover the first time such a configuration is reached, the actions taken by $\pda$ cannot depend on $w$ since at no point was $w$ exposed at the top of the stack. Hence it must be the case that $pA \Rightarrow q$.

The $\pda$-automaton when accepting a stack content $A_{1} \ldots A_{n} \bot$ from
the state $p$ will guess the states $p_{1}, \ldots, p_{n}$ such that $pA_{1} \eRb{}{\pda} p_{1}$ and $p_{i}A_{i+1} \eRb{}{\pda} p_{i+1}$ for $i \in [0,n-1]$ and will enter a final state upon reading the symbol $\bot$ if $p_{n}\bot \eRb{}{\pda} q_{f}\bot$.

Consider the $\pda$-automaton $\paut$ with set of states $Q \cup \{s_{\bot}\}$ where $s_{\bot}$ is a new state and the only final state of the automaton.
The transitions of the automaton $\paut$ are defined as follows:
\begin{itemize}
\item $p \era{A} q$ if and only if $pA \eRb{}{\pda} q$ for all $p,q \in Q$ and $A \in \Gamma \setminus \{\bot\}$,
\item $p \era{\bot} s_{\bot}$ if and only if $p\bot \eRb{}{\pda} q_{f}\bot$ for all $p \in Q$.
\end{itemize}

A simple induction on the length of the stack content shows that $\paut$
accepts a stack content $w\bot$ from the state $q \in Q$ if and only if
$(q,w\bot)$ belongs to $\prestar{}{\{(q_{f},\bot)\}}$.

To make the construction effective, it remains to  compute
the relations $pA \Rightarrow q$ and $p\bot \Rightarrow q\bot$ for all states $p$ and $q \in Q$ and stack symbol $A \in \Gamma$. The procedure provided
by \buchi is exponential\footnote{In \cite{Buchi64}, the $\pda$-automaton constructed is deterministic (essentially the automaton obtained by applying the power-set construction to the automaton presented here). With the added constraint of determinism, it not possible to obtain a polynomial algorithm as the smallest deterministic automaton is in general exponential in the size of the pushdown system. To convince oneself, it is enough to consider a pushdown system that simulates a non-deterministic finite state automaton (NFA) by popping its stack until the bottom of the stack is reached and when the bottom of the stack is reached goes to the state $q_{f}$ if the NFA has reached a final state.}. He first establishes a bound on the height of the stack necessary to build a derivation path witnessing these relations. As the bound is polynomial in the size of the pushdown system, the problem is reduced to a simple reachability problem in a finite graph of exponential size with respect to the size of the pushdown system.

\newcommand{\Rew}{\mathrm{Rew}}

To obtain a polynomial algorithm, it is enough to efficiently compute the relation $Rew = \{ (pA,qB) \mid pA \eRb{}{\pda} qB \}$. Indeed $pA \eRb{}{\pda} q$ if and only
if there exists $r \in Q$ and $B \in \Gamma$ such that $pA \eRb{}{\pda} rB$
(\emph{i.e.} $(p,A,r,B) \in \Rew$) and $rB \rightarrow q$ is a transition of $\pda$.

The key idea which is at the heart\footnote{We will see that the algorithm presented in Section~\ref{ssec:saturation} performs a fixed-point computation for the relation $\{ (pA,q) \mid pA \eRb{}{\pda} q \}$.} of the saturation algorithm presented in Section~\ref{ssec:saturation} is to express $\Rew$ as a smallest fixed-point.

The relation $\Rew$ is the smallest relation (for the inclusion) in $Q\Gamma \times Q \Gamma$ such that:
\begin{itemize}
\item $(pA,pA) \in Rew$ for all $p \in Q$ and $A \in \Gamma$,
\item $(pA,qB) \in \Rew$ if $pA \rightarrow qB$ is a transition of $\pda$,
\item $(pA,qC) \in \Rew$ if $(pA,rB) \in \Rew$ and $(rB,qC) \in \Rew$,
\item $(pA,qC) \in \Rew$ if $pA \rightarrow rBC$ is a transition of $\pda$ and there exists $t \in Q$ and $D \in \Gamma$ such that $(rB,tD) \in \Rew$ and $tD \rightarrow q$ is a transition of $\pda$.
\end{itemize}

The property $(1)$ expresses that $\Rew$ is reflexive and $(3)$ that it is transitive. Property $(2)$ ensures that $\Rew$ contains the relevant transitions of $\pda$. Property $(4)$ describes the case when $pA \eRb{}{\pda} qC$ is obtained by a sequence of the form $pA \erb{}{\pda} rBC \eRb{}{\pda} qC$ where
$rB \eRb{}{\pda} q$.

Using the Knaster-Tarski theorem, we can compute $\Rew$ as the limit of an increasing sequence of relations $(\Rew_{i})_{i \geq 0}$ over $Q \times \Gamma$. The relation $\Rew_{0}$ contains the elements satisfying property $(1)$ and $(2)$. The relation $\Rew_{i+1}$ is obtained from $\Rew_{i}$ by adding all the elements that can be derived by property $(3)$ or $(4)$ in $\Rew_{i}$. The sequence $(\Rew_{i})_{i \geq 0)}$ is increasing for the inclusion and its limit (\emph{i.e.} the first set such that $\Rew_{i+1}=\Rew_{i}$) is equal to $\Rew$. As at least one element is added at each step before the limit is reached, the limit is reached in at most $|Q|^2 |\Gamma|^{2}$ steps. Furthermore as the computation of $\Rew_{i+1}$ from $\Rew_{i}$ can be done in polynomial time with respect to the size of $\pda$, the resulting algorithm is polynomial. However the exact complexity is not as good as the algorithm presented in Section~\ref{ssec:saturation}.

\subsection{Saturation algorithm of \cite{BEM97}}
\label{ssec:saturation}

In \cite{BEM97}, Bouajjani \emph{et al.} present an algorithm  that given
a pushdown system $\pda=(Q,\Gamma,\bot,\Delta)$ and  a $\pda$-automaton $\paut=(\astates,Q,\adelta,\afinals)$, constructs a new $\pda$-automaton $\pautb$ accepting $\prestar{\pda}{\langof{\paut}}$.
The only requirement on $\paut$ is that no transition in $\adelta$ goes
back to a state in $Q$\footnote{This requirement is easily met by adding a copy of each state in $Q$ if necessary. This restriction is required to ensure that the first invariant maintained by the algorithm holds initially.}. This restriction also implies that none of the states in $Q$ are final.

The algorithm proceeds by adding transitions to $\paut$ following a unique rule until no new transition can be added. The resulting $\pda$-automaton $\pautb$ accepts the set of configurations $\prestar{\pda}{\langof{\paut}}$.

\noindent
More precisely, the algorithm constructs a finite sequence $(\paut_{i})_{i \in [0,N]}$ of $\pda$-automa\-ta. The  $\pda$-automaton $\paut_{0}$ is the automaton $\paut$. All the $\pda$-automata  $\paut_{i}$ are of the form $(\astates,Q,\afinals,\adelta_{i})$, meaning that they only differ by their set of transitions. The construction guaranties that for all $i \in [0,N-1]$, $\adelta_{i} \subseteq \adelta_{i+1}$ and terminates when $\adelta_{i+1}=\adelta_{i}$. As at least one transition is added at each step, the algorithm terminates in at most $|Q|^{2} |\Gamma|$ steps.

The set of transitions $\adelta_{i+1}$ is obtained by adding to $\adelta_{i}$, the transition:
\begin{center}
$ p \erb{A}{} s$ if $q \eRb{w}{\paut_{i}} s$ and $pA \rightarrow qw$ is a transition of $\pda$.
\end{center}
Note that only transitions starting with a state of $Q$ are added by the algorithm. In particular, the language accepted the automaton $\paut_{i}$ from
a state in $\astates \setminus Q$ never changes.\footnote{Recall that initially the states in $Q$ are not the target of any transition}.

The construction of $\adelta_{i+1}$ from $\adelta_{i}$ ensures that the configurations that can reach in one step a configuration in $\langof{\paut_{i}}$ belong
to $\langof{\paut_{i+1}}$. Consider two configurations  $c=(p,Au)$ and $c'=(q,wu)$ such that $pA \rightarrow qw$ is a transition of $\pda$ (and hence $c \erb{}{\pda} c'$).
Now assume that $c'$ belongs to $\langof{\paut_{i}}$. This means that for some state $s \in \astates$ and some final state $s_{f} \in \afinals$, $q \eRb{w}{\paut_{i}} s \eRb{u}{\paut_{i}} s_{f}$. The rule of construction of $\adelta_{i+1}$ ensures that
$p \erb{A}{} s$ is a transition of $\paut_{i+1}$. Hence $p \eRb{A}{\paut_{i+1}} s \eRb{u}{\paut_{i+1}} s_{f}$ and the configuration $c=(p,Au)$ is accepted by $\paut_{i+1}$. As $\pautb$ is the limit of the saturation process (\emph{i.e.} $\pautb=\paut_{N-1}=\paut_{N}$), $\langof{\pautb}$ is closed under taking the 
immediate predecessor for the relation $\erb{}{\pda}$ (\emph{i.e.} if $c' \in \langof{\pautb}$ and 
$c \erb{}{\pda} c'$ then $c \in \langof{\pautb}$). As $\langof{\pautb}$ includes $\langof{\paut}$, it follows that $\prestar{\pda}{\langof{\paut}} \subseteq \langof{\pautb}$.

The proof of the converse inclusion requires a more careful analysis. The algorithm maintains two invariants on the transitions in $\adelta_{i}$. For all $i \in [0,N]$,  the presence of a transition $p \erb{A}{} s$ in $\adelta_{i}$ guaranties that:
\begin{enumerate}
\item  $pA \eRb{}{\pda} s$ if $s$ belongs to $Q$.
\item  the configuration $(p,Au)$ belongs to $\prestar{}{\langof{\paut}}$
for any $u \in \slang{s}{\paut_{i}}=\slang{s}{\paut}$ if $s$ belongs to $\astates \setminus Q$.
\end{enumerate}

From these invariants, it follows that for all $i \geq 0$, $\langof{\paut_{i}} \subseteq \prestar{\pda}{\langof{\paut}}$. In particular, $\langof{\pautb} \subseteq \prestar{\pda}{\langof{\paut}}$.

\begin{remark}
As indicated by the first invariant, if we restrict our attention to transitions with both source and target in $Q$, this algorithm is
performing a fixed-point computation for the relation $\eRb{}{\pda}$ restricted
to $(Q \times \Gamma) \times (Q \times \{\varepsilon\})$.
Indeed this relation can be characterised as the smallest relation  (for the inclusion) $\mathcal{R}$  such that:
\begin{enumerate}
\item $pA \,\mathcal{R}\, q$ if $pA \rightarrow q$ belongs to $\Delta$,
\item $pA \,\mathcal{R}\, q$ if $rB \,\mathcal{R}\, q$  and $pA \rightarrow rB$ belongs to $\Delta$,
\item $pA \mathcal{R} q$ if $pA \rightarrow rBC$ belongs to $\Delta$ and for some state $s \in Q$, $rB \mathcal{R} s$ and $sC \mathcal{R} q$.
\end{enumerate}
In fact, the algorithm performs the computation of the smallest such relation following the procedure given by Knaster-Tarski theorem.
\end{remark}

A naive implementation of this algorithm yields a complexity in $\mathcal{O}(|\pda|^{2}|\paut|^{3})$. However a more efficient implementation
presented in \cite{EHRS00} lowers the complexity to $\mathcal{O}(|Q|^{2}|\Delta|)$.

In \cite{EHRS00}, an adaptation of the algorithm for computing $Pre^{*}$
is given to compute $Post^{*}$. The algorithm is slightly less elegant as it requires the addition of new states before the saturation process. In fact, it is very similar to first applying the transformation to invert the pushdown system presented at the beginning of this section and then applying the algorithm to compute $Pre^{*}$.

In \cite{S02}, Schwoon shows how to use the saturation algorithm to construct for any configuration $c$ accepted by $\pautb$ a derivation path
to some configuration in $\langof{\paut}$.

\subsection{Derivation relation of a pushdown system}
\label{ssec:derivation}

\newcommand{\Deriv}{\mathrm{Deriv}}
\newcommand{\Behaviour}{\mathrm{Behaviour}}
\newcommand{\Red}{\mathrm{Red}}

In this section, we will see that the saturation method can be adapted
to characterise the derivation relation of a pushdown system. Let us fix
a pushdown system\footnote{To simplify the presentation, we do not take the bottom of stack symbol into account.} $\pda=(Q,\Gamma,\Delta)$, an initial state $q_{0}$ and a final state $q_{f}$. We aim at giving an effective characterisation of the following relation between stacks:
\[
\Deriv_{\pda} =\{ (u,v) \in \Gamma^{*} \mid (q_{0},u) \eRb{}{\pda} (q_{f},v) \}.
\]

In \cite{Caucal88}, Caucal showed that $\Deriv_{\pda} \subseteq \Gamma^{*} \times \Gamma^{*}$ is a rational relation, \emph{i.e.} it is accepted by a finite state automaton with output (also called a transducer).

The proof presented here is based on \cite{Caucal08} but similar ideas
can be found in \cite{Sakarovitch09,FWW97}. The idea of the proof is to use symbols to represent the actions  of the pushdown system on the stack: one symbol for pushing a given symbol and one symbol for popping it. The pushdown system is transformed into a finite state automaton that instead of performing the actions on the stack  outputs the symbols that represent these actions (see Section~\ref{sssec:actions}). This finite state automaton is then transformed using a saturation algorithm so that it erases sequences of  actions corresponding to pushing a symbol and then immediately popping it (see Section~\ref{sssec:saturation}). From this \emph{reduced} language, the relation $\Deriv_{\pda}$ is easily characterised (see Section~\ref{sssec:characterization}).

\subsubsection{Sequences of  stacks actions}
\label{sssec:actions}

 For every symbol $A \in \Gamma$, we introduce two symbols:
\begin{itemize}
\item
 $A_{+}$ which represents the action of pushing the symbol $A$ on top of the stack,
\item and $A_{-}$ which represents the action of popping the symbol $A$ from the top of the stack.
\end{itemize}
We denote  by $\Gamma_{+}$ the set $\{ A_{+} \mid A \in \Gamma\}$ of \emph{push} actions, by $\Gamma_{-}$ the set $\{ A_{-} \mid A \in \Gamma\}$ of \emph{pop} actions and by $\overline{\Gamma}$ the set $\Gamma_{+} \cup \Gamma_{-}$ of all action symbols.

Intuitively a sequence $\alpha=\alpha_{1} \ldots \alpha_{n} \in \overline{\Gamma}^{*}$ is interpreted as performing  the action $\alpha_{1}$, followed by the action $\alpha_{2}$ and so on. For instance, the effect on the stack of the transition $pA \rightarrow qBC$ is represented by the word $A_{-}C_{+}B_{+}$. First the automaton removes the $A$ from the top of the stack and then pushes $C$ and then $B$.

For two stacks $u$ and $v \in \Gamma^{*}$, we write $u \eVb{\alpha}{} v$ if $u$ can be transformed into $v$ by the sequence of actions $\alpha$. For instance, we have $ABB \eVb{\alpha}{} DCB$ for the $\alpha$ sequence $A_{-}B_{-}C_{+}D_{+}$. Note that some sequences of actions such as $B_{+} C_{-}$ cannot be applied to any stack. We say that such sequences $\alpha$ are \emph{non-productive}, \emph{i.e.}  there are no $u$ and $v \in \Gamma^{*}$ such that $u \eVb{\alpha}{} v$.

From the pushdown system $\pda$, we can construct a regular set of action sequences denoted $\Behaviour_{\pda}$ which contains all the sequences (even the non-productive ones) that can be performed by $\pda$ when starting in state $q_{0}$
and ending in state $q_{f}$. Consider for instance the finite state automaton\footnote{The finite state automaton does not strictly conform to the definition we gave in Section~\ref{sec:preliminaries} as its transitions are labelled by words and not single letters. This can be easily avoided at the cost of adding intermediate states.} $(Q,\set{q_{0}},\set{q_{f}},\adelta)$ where the set of transitions $\adelta$ is given by:
\[
\left\{
\begin{array}{lcl}
    p \erb{A_{-}C_{+}B_{+}}{} q \in \adelta \quad \text{if} \quad pA \rightarrow qBC \in \Delta \\
    p \erb{A_{-}B_{+}}{} q \in \adelta \quad \text{if} \quad pA \rightarrow qB \in \Delta \\
 p \erb{A_{-}}{} q \in \adelta \quad \text{if} \quad pA \rightarrow q \in \Delta \\
\end{array}
\right.
\]

It is clear that $\Behaviour_{\pda}$ characterises $\Deriv_{\pda}$ in the following sense:
\[
(u,v) \in \Deriv_{\pda} \quad \text{if and only if} \quad u \eVb{\alpha}{} v \;\;\text{for some}\; \alpha \in \Behaviour_{\pda}.
\]

However this representation of $\Deriv_{\pda}$ is not yet very helpful. For instance, $\Behaviour_{\pda}$ can contain non-productive sequences or
sequences such as $A_{-} B_{+} A_{+} A_{-} C_{+} C_{-}$ which is equivalent to the more informative sequence $A_{-} B_{+}$.

\subsubsection{Reducing  sequences of actions}
\label{sssec:saturation}

To simplify  $\Behaviour_{\pda}$, we first erase all factors of the form $A_{+}A_{-}$ for $A \in \Gamma$. These factors can safely be omitted as they do not affect the stack: the symbol is pushed then immediately popped. A sequence that does not contain any such factors is called \emph{reduced}.

To perform this erasure, we introduce the relation $\mapsto$ which relates a
stack $u \in \Gamma^{*}$ and a stack $v \in \Gamma^{*}$ if $v$ can be obtained by erasing a factor $A_{+}A_{-}$ from $u$ (\emph{i.e.} $u=u_{1}A_{+}A_{-}u_{2}$ and $v=u_{1}u_{2}$). Clearly, if $\alpha \mapsto \beta$ then the sequences $\alpha$ and $\beta$ are equivalent with respect to their actions on the stack :
\begin{center}
for $u,v \in \Gamma^{*}$, $u \eVb{\alpha}{} v$ if and only if $u \eVb{\beta}{} v$.
\end{center}

As the rewriting relation $\mapsto$ is confluent and decreases the length of the sequence, every sequence $\alpha$ can be iteratively rewritten by $\mapsto$ into a reduced sequence denoted $\Red(\alpha)$. For instance the reduced sequence
associated to $B_{-}A_{+}A_{+}A_{-}A_{-}C_{+}$ is $B_{-}C_{+}$ as $B_{-}A_{+}A_{+}A_{-}A_{-}C_{+} \mapsto B_{-}A_{+}A_{-}C_{+} \mapsto B_{-}C_{+}$.

In \cite{Benois69}, Benois showed\footnote{Benois consider the erasure of all factor of the form $A_{-}A_{+}$ as well as $A_{+}A_{-}$ but the proof is identical.}
that the set of reduced sequences corresponding to a regular set of sequences is again regular.

\begin{theorem}{\cite{Benois69,Benois86}}
For any regular set $R$ of action sequences, the corresponding set of reduced action sequences:
\[
\Red(R) = \{ \Red(\alpha)  \mid \alpha \in R \}
\]
\noindent
is regular. Moreover given a finite automaton $\mathcal{A}$ accepting $R$, an automaton accepting $\Red(R)$ can be constructed in $\mathcal{O}(|\mathcal{A}|^{3})$.
\end{theorem}

The proof of this theorem is the essence of the saturation method. Starting with the automaton $\mathcal{A}$, $\varepsilon$-transitions are added until no new $\varepsilon$-transition can be added. The $\varepsilon$-transitions are added according to the following rule. We add an $\varepsilon$-transition from a state $p$ to a state $q$ if it is possible to reach $q$ from $p$ reading a word of form  $A_{+} \varepsilon^{*} A_{-}$.
It can be shown that the resulting saturated automaton accepts the language:
\[
\{ \beta \in \overline{\Gamma}^{*} \mid \alpha \mapsto^{*} \beta \quad\text{for some $\alpha \in R$} \}.
\]
The construction is concluded by taking the $\varepsilon$-closure of the saturated automaton  and restricting the language to the set of reduced sequences (which is a regular language as it is the complement of the language $\cup_{A \in \Gamma}\overline{\Gamma}^{*}A_{+}A_{-}\overline{\Gamma}^{*}$). A careful implementation of the procedure presented in \cite{Benois86} gives an algorithm in $\mathcal{O}(|\mathcal{A}|^{3})$.

\subsubsection{Characterisation of $\Deriv_{\pda}$}
\label{sssec:characterization}

One of the advantages of working with $\Red(\Behaviour_{\pda})$ is that we can easily remove non-productive sequences. Indeed a reduced sequence is non-productive
if and only if it contains a factor of the form $A_{+} B_{-}$ for $A \neq B \in \Gamma$.

We can hence compute the regular language:
\[
  RP_{\pda} = \Red(\Behaviour_{\pda}) \cap \left(\overline{\Gamma}^{*} \setminus \bigcup_{A \neq B \in \Gamma} \overline{\Gamma}^{*}A_{+}B_{-}\overline{\Gamma}^{*}\right)
\]
which is composed of  the reduced and productive action sequences characterising $\Deriv_{\pda}$.

The language $RP_{\pda}$ does not contain any factor in $\Gamma_{+}\Gamma_{-}$ and is hence included in $\Gamma_{-}^{*}\Gamma_{+}^{*}$. We can express it as a finite union:
\[
   \bigcup_{i \in [1,N]} X_{i}Y_{i}
\]
where for all $i \in [1,N]$, $X_{i}$ is a regular language in $\Gamma_{-}^{*}$ and
$Y_{i}$ is a regular language in $\Gamma_{+}^{*}$.

Let us denote by $U_{i}$ the regular set $\{ A^{1} \cdots A^{n} \in \Gamma^{*} | \mid A^{1}_{-} \cdots A^{n}_{-} \in X_{i} \}$ of words in $\Gamma^{*}$ that can be popped by a sequence in $X_{i}$ and by $V_{i}$ the regular set $\{ A^{1} \cdots A^{n} \in \Gamma^{*} | \mid A^{n}_{+} \cdots A^{1}_{+} \in Y_{i} \}$ of words in $\Gamma^{*}$ that can be pushed by a sequence in $Y_{i}$. 

The relation $\Deriv_{\pda}$ can be characterised as follows: a pair $(w_{1},w_{2})$ belongs  to $\Deriv_{\pda}$, if for some $i \in [1,N]$, $w_{1}$ can be written as $uw$ with $u \in U_{i}$ and $w_{2}$ can be written as $vw$ for some $v \in V_i$. In other terms, the relation $\Deriv_{\pda}$ can be written as a finite union of relations that remove a prefix of the stack belonging to a certain regular language and replace any word in another regular language. As these relations are easily accepted by finite transducer, so is $\Deriv_{\pda}$. Combining all the steps, we obtain a polynomial time algorithm for computing a transducer accepting $\Deriv_{\pda}$ from $\pda$.


%% file: games.tex
\newcommand\altA{\mathcal{A}}
\newcommand\altB{\mathcal{B}}
\newcommand\infocc{\mathrm{Inf}}
\newcommand\gcoli{\alpha}
\newcommand\gcolj{\beta}
\newcommand\gcolk{\gamma}
\newcommand\gmaxcol{\kappa}
\newcommand\minf{\mathrm{min}}
\newcommand\prestep[2]{Pre_{#1}({#2})}
\newcommand\prestepc[3]{Pre^{#2}_{#1}({#3})}
\newcommand\qend{s_\bot}
\newcommand\qall{s^\ast}
\newcommand\pinit[2]{#1^#2}
\newcommand\psatinitA{\altA^0}
\newcommand\fpstep{Fix}
\newcommand\fpstepb[1]{Fix$_{#1}$}
\newcommand\fppre{Pre}
\newcommand\fpproj[1]{Proj}
\newcommand\proj[2]{\pi_{#1, #2}}
\newcommand\fpdispatch{Dispatch}
\newcommand\fpstepp{Fix}

\section{Winning regions of pushdown games}
\label{sec:games}

The saturation technique also generalises to the analysis of pushdown games with
two players: \eloise and \abelard.  The two players may, for example, model a
program (\eloise) interacting with the environment (\abelard).  While the
program can control its next move based on its internal state, it cannot control
the results of requesting external input.  Hence, the external input is decided
by the second player.

A pushdown game may be used to analyse various types of properties.  We will
consider three, increasingly expressive, types of properties here: reachability,
\buchi and parity.  We will begin by defining games with generic winning
conditions and then consider the instantiations of this generic framework for
each winning condition in turn.  We will simultaneously discuss the saturation
algorithm for each of these properties and show how they build upon each other.

The saturation algorithm was first extended to pushdown reachability games by
Bouajjani\etal~\cite{BEM97}.  Their algorithm was extended to the case of \buchi
games by Cachat~\cite{Cachat02} and then to parity games by Hague and
Ong~\cite{HagueO09}.  Our presentation will follow that of Hague and Ong since
it provides the most general algorithm, though we remark that all the essential
ideas of the algorithm were in place by the introduction of the \buchi
algorithm.  The main contribution of Hague and Ong was a proof framework that
simplified the technical arguments by Bouajjani\etal and Cachat and allowed the
full parity case to go through.

\subsection{Preliminaries}

\subsubsection{Pushdown games}

We can obtain a two-player game from a pushdown system $\pda$ by the addition of
two components: a partition of the configurations of $\pda$ into positions
controlled by \eloise and positions controlled by \abelard; and the definition
of a winning condition that determines the winner of any given play of the game.

In the following, for technical convenience, we will assume for each $q \in Q$
and $A \in \Gamma$ there exists some $(q, A) \rightarrow (p, w) \in \Delta$.
Together with the bottom-of-stack symbol, this condition ensures that from a
configuration $(q, w\bot)$ it is not possible for the system to become stuck;
that is, reach a configuration with no successor.

A two-player pushdown game is a tuple $\pda = (Q, \Gamma, \bot, \Delta,
\wincond)$ such that $(Q, \Gamma, \bot, \Delta)$ defines a pushdown system,
$Q$ is partitioned $Q = Q_E \uplus Q_A$ into
\eloise and \abelard positions respectively, and $\wincond$ is a set of infinite
sequences of configurations of $\pda$.

A play of a pushdown game is an infinite sequence $(q_0, w_0), (q_1, w_1),
\ldots$ where $(q_0, w_0)$ is some starting configuration and $(q_{i+1},
w_{i+1})$ is obtained from $(q_i, w_i)$ via some transition $(q_i, A)
\rightarrow (q_{i+1}, w) \in \Delta$.  In the case where $q_i \in Q_E$ it
is \eloise who chooses the transition to apply, otherwise \abelard chooses the
transition.

The winner of an infinite play $(q_0, w_0), (q_1, w_1), \ldots$ is \eloise if
$(q_0, w_0), (q_1, w_1), \ldots \in \wincond$; otherwise, \abelard wins the
play.  The winning region $\winningregion$ of a pushdown game is the set of all
configurations from which \eloise can always win all plays, regardless of the
transitions chosen by \abelard.

\subsubsection{Alternating automata}

To extend the saturation algorithm to compute the winning region of a pushdown
game, we augment the automata used to recognise sets of configurations with
alternation.  Bouajjani\etal first used alternating automata to analyse pushdown
reachability games via saturation~\cite{BEM97}, however, they used the
equivalent formalism of \emph{alternating pushdown systems} rather than pushdown
games.  An alternating automaton is a tuple $\altA = (\astates, \Gamma,
\afinals, \adelta)$ where $\astates$ is a finite set of states, $\Gamma$ is a
finite alphabet, $\afinals \subseteq \astates$ is the set of accepting states,
and $\adelta \subseteq \astates \times \Gamma \times 2^\astates$ is a transition
relation.  Note that we do not specify a set of initial states.  This is because
it is more convenient to present the following results in terms of the stacks
accepted from particular states, rather than fixing a set of initial states.

Whereas a transition $s \xrightarrow{A} t$ of a non-deterministic automaton
requires the remainder of the word to be accepted from $t$, a transition $s
\xrightarrow{A} S$ of an alternating automaton requires that the remainder of
the word is accepted from all states $s' \in S$.  It is this ``for all''
condition that captures the fact that \eloise must be able to win for all moves
\abelard may make.

More formally, a run over a word $A_1\ldots A_n \in \Gamma^\ast$ from a state
$s_0$ is a sequence
\[
    S_1 \xrightarrow{A_1} \cdots \xrightarrow{A_n} S_{n+1}
\]
where each $S_i$ is a set of states such that $S_1 = \set{s_0}$, and for each $1
\leq i \leq n$ we have
\[
    \forall s \in S_i . \exists s \xrightarrow{A_i} S \in \adelta \wedge S
    \subseteq S_{i+1} \ .
\]
The run is accepting if $S_{n+1} \subseteq \afinals$.  Thus, for a given state
$s$, we define $\slang{s}{\altA}$ to be the set of words over which there is an
accepting run of $\altA$ from $\set{s}$.

When $S_i$ is a singleton set, we will often omit the set notation.  For
example, the run above could be written
\[
    s_0 \xrightarrow{A_1} \cdots \xrightarrow{A_n} S_{n+1} \ .
\]
Further more, when $w = A_1\ldots A_n$ we will write $s \xrightarrow{w}
S$ as shorthand for a run from $s$ to $S$.

\subsection{Pushdown reachability games}

One of the simplest winning conditions for a game is the reachability condition.
Given a target set of configurations $C$, the reachability condition states that
\eloise wins the game from a given configuration if she can force all plays
starting at that configuration to some configuration in $C$.

That is, a pushdown reachability game is a tuple $(Q, \Gamma, \bot, \Delta, C)$
such that $(Q, \Gamma, \Delta, \wincond)$ is a pushdown game where
\[
    \wincond = \setcomp{c_0, c_1, \ldots}{\exists i . c_i \in C}
\]
is the set of all sequences of configurations containing some configuration in
$C$.

\subsubsection{Characterising the winning region}

In the sequel we will need to combine least and greatest fixed points.  We will
use $\mu$ to denote the least fixed point operator, and $\nu$ to denote the
greatest fixed point operator.

In the simple case of reachability for a pushdown system $\pda$ and set of
target configurations $C$ we can characterise the winning region $\winningregion
= \prestar{\pda}{C}$ as
\[
    \mu Z . C \cup \prestep{\pda}{Z}
\]
where
\[
    \prestep{\pda}{Z} = \setcomp{(p, w)}{\begin{array}{rcll}
        p \in Q_E &\Rightarrow& \exists (p, w) \rightarrow c .\; c \in Z  & \land \\
        p \in Q_A &\Rightarrow& \forall (p, w) \rightarrow c .\; c \in Z
    \end{array}} \ .
\]
That is, to appear in $\winningregion$ for a configuration belonging to \eloise,
it must be possible for her to choose a transition that progresses towards $C$.
For configurations belonging to \abelard, it must be the case that he cannot
help but choose a transition that progresses towards $C$.

\subsubsection{Computing the winning region}

Fix a pushdown reachability game $\pda = (Q, \Gamma, \Delta, C)$.  We will show
how to construct an automaton $\altB$ whose state set includes the state $p$
for all $p \in Q$ and $w \in \slang{p}{\altB}$ iff $(p, w) \in \winningregion$.

Computing \eloise's winning region is a direct extension of the saturation
algorithm for $\prestar{\pda}{C}$ in the non-game setting.  We assume $C$ is a
regular set of configurations represented by an alternating automaton $\altA =
(\astates, \Gamma, \adelta, \afinals)$ such that $Q \subseteq \astates$ and there
are no-incoming transitions to any state in $Q$.

The saturation algorithm constructs the automaton $\altB$ that is the least
fixed point of the sequence of automata $\altA_0, \altA_1, \ldots$ where
$\altA_0 = \altA = (\astates, \Gamma, \adelta_0, \afinals)$ and $\altA_{i+1} =
(\astates, \Gamma, \adelta_{i+1}, \afinals)$ where $\adelta_{i+1}$ is the
smallest set of transitions such that
\begin{enumerate}
    \item $\adelta_i \subseteq \adelta_{i+1}$, and

    \item for each $q \in Q_E$, if $(q, A) \rightarrow (p, w) \in \Delta$ and $p
        \xrightarrow{w} S$ is a run of $\altA_i$, then
        \[
            q \xrightarrow{a} S \in \adelta_{i+1}
        \]
        and

    \item for each $q \in Q_A$ and $A \in \Gamma$ and $S \subseteq \astates$
        such that for all
        \[
            (q, A) \rightarrow (p, w) \in \Delta
        \]
        there exists a run $p \xrightarrow{w} S'$ of $\altA_i$ with $S'
        \subseteq S$, we have
        \[
            q \xrightarrow{a} S \in \adelta_{i+1} \ .
        \]
\end{enumerate}
One can prove that $(p, w) \in \winningregion$ iff $w \in \slang{p}{\altB}$.
Thus we obtain regularity of the winning region.  Since the maximum number of
transitions of an alternating automaton is exponential in the number of states
(and we do not add any new states), we have that $\altB$ is constructible in
exponential time.

\begin{theorem}
    The winning region of a pushdown reachability game is regular and constructible in
    exponential time.
\end{theorem}

\subsubsection{Winning strategies}

Cachat has given two realisations of \eloise's winning strategy in a pushdown
reachability game from a configuration in her winning region~\cite{Cachat02} .
The first is a positional strategy that requires space linear in the size of the
stack to compute.  That is, he gives an algorithm that reads the stack and
prescribes the next move that \eloise should make in order to win the game.  The
algorithm assigns costs to accepting runs of $\altB$ for configurations in
$\winningregion$ by summing costs assigned to individual transitions.

Alternatively, Cachat presents a strategy that can be implemented by a pushdown
automaton that tracks the moves of \abelard and recommends moves to \eloise.
Since the automaton tracks the game, the strategy is not positional.  However,
the prescription of the next move requires only constant time.

In his PhD. thesis~\cite{Cachat03}, Cachat also argues that similar strategies
can be computed for \abelard for positions in his winning region.

\subsection{Pushdown \buchi games}

Plays of a game are infinite sequences.  The reachability condition only depends
on finite prefixes of these plays, hence games are won  within a finite
number of moves.  This prevents the specification of liveness properties such as
``every request is followed by an acknowledgment''.  Since it is not possible to
know when to ``stop waiting'' for an acknowledgment to arrive, it is not
possible to specify such conditions as simple reachability properties.

\buchi conditions allow liveness properties to be defined since deciding the
winner of a particular play can take the whole infinite sequence into account.
We define a pushdown \buchi game as a tuple $(Q, \Gamma, \bot, \Delta, F)$ --
where $F \subseteq Q$ is a set of target control states -- which defines a
pushdown game $(Q, \Gamma, \bot, \Delta, \wincond)$ with
\[
    \wincond = \setcomp{(p_0, w_0), (p_1, w_1), \ldots}{\forall i . \exists j
    \geq i . p_j \in F} \ .
\]
That is, \eloise wins the play if there is some control state in $F$ that is
visited infinitely often.

Cachat generalised the saturation method to construct the winning region of a
pushdown \buchi game~\cite{Cachat02} by introducing the nesting of fixed point
computations and projection described below.

To characterise the winning region of a pushdown \buchi game, a single least
fixed point computation no longer suffices.  Intuitively this is because
satisfying the \buchi condition amounts to repeatedly satisfying a reachability
condition; that is, repeatedly reaching a control state in $F$.  We will begin
by giving the characterisation, and then decoding it in the following
paragraphs.  By abuse of notation, we will write $F$ to also denote the set of
configurations $\setcomp{(p, w)}{p \in F}$ and $\negation{F}$ to denote its
complement.  The winning region of \eloise can be defined as
\[
    \nu Z_0 . \mu Z_1 . \brac{F \cap \prestep{\pda}{Z_0}} \cup
    \brac{\negation{F} \cap \prestep{\pda}{Z_1}} \ .
\]
There are two pre-steps in the formula: $\prestep{\pda}{Z_0}$ and
$\prestep{\pda}{Z_1}$.  When a configuration is in $F$ then we require that
\eloise can force the next step of play to stay within $Z_0$.  When the
configuration is not in $F$ we require that \eloise can force play to stay
within $Z_1$.

To understand the role of the different fixed points, imagine a game where there
is only one move from some configuration $(p, w)$
\[
    (p, w) \rightarrow (p, w) \ .
\]
In the case where $p \in F$ it will be the case that $(p, w)$ appears in the
greatest fixed point $Z_0$.  This is because greatest fixed points can be
``self-supporting'': if we include $(p, w)$ in an approximation of $Z_0$, then
it will appear in the next approximation of $Z_0$ by virtue of the fact that it
was in the old valuation.

In the other case, when $p \notin F$, we would require $(p, w)$ to appear in the
least fixed point $Z_1$.  However, since the least fixed point is the smallest
possible fixed point, its members cannot be self-supporting.  That is, if we
took $(p, w)$ out of our approximation, the next approximation would not include
$(p, w)$: there is nothing external compelling $(p, w)$ to be in the least fixed
point.  This is why a reachability property is a least fixed point: it must
contain only the configurations that eventually reach a target configuration --
it cannot put off satisfying this obligation for an infinite number of steps.

In terms of \buchi games this difference makes sense: a play that repeatedly
visits only the configuration $(p, w)$ is only winning if $p \in F$.  If $p
\notin F$ then a configuration can only be winning if it eventually (after a
finite number of steps) moves to a configuration that has a control state in
$F$.  Thus, the least fixed point represents configurations that must eventually
reach a ``good'' configuration, while the greatest fixed point represents good
configurations that are able to support themselves.

\subsubsection{Computing the winning region}

\paragraph{Automaton representation of multiple fixed points}

The saturation method for reachability properties computed a single fixed point
with a single fixed point variable.  We can think of the successive automata
$\altA_0, \altA_1, \ldots$ as successive approximations of the value of $Z$.
The final automaton computed gives the value of $Z$ that is the solution to
\[
    \mu Z . C \cup \prestep{\pda}{Z} \ .
\]

In the case of \buchi games, there are two nested fixed point computations over
the variables $Z_0$ and $Z_1$.  The winning region is the greatest fixed point
for $Z_0$.  However, in order to compute this fixed point we also have to
compute the least fixed point for $Z_1$.  Hence, we will need an automaton that
can represent two different sets of configurations: the approximation of $Z_0$
as well as the approximation of $Z_1$.  Thus, instead of having a state $p$ of
the alternating automaton for each control state $p$, we will have two states
$\pinit{p}{0}$ and $\pinit{p}{1}$.  A configuration $(p, w)$ appears in the
current approximation of $Z_0$ if it is accepted from $\pinit{p}{0}$, and it
appears in the current approximation of $Z_1$ if it is accepted from
$\pinit{p}{1}$.  We will also use control states of the form $\pinit{p}{2}$ to
hold intermediate values of the computation.

Finally, the automata we build will have two additional states (these will be
the only states that are not of the form $\pinit{p}{\gcoli}$ for some $\gcoli$).  There
will be one state $\qend$ that will be the only accepting state.  Since all
stacks finish with the bottom-of-stack symbol $\bot$, this state will have no
outgoing transitions, and all incoming transitions will be of the form $s
\xrightarrow{\bot} \set{\qend}$.  No other transitions in the automaton will be
labelled $\bot$.

The other additional state is $\qall$ from which all stacks are accepted.  This
state has the outgoing transitions $\qall \xrightarrow{A} \set{\qall}$ for all
$A \in \Gamma$ with $A \neq \bot$, and $\qall \xrightarrow{\bot} \set{\qend}$.

\paragraph{Evaluation strategy}

The saturation method computes fixed points following Knaster-Tarski theorem.
  That is, to compute a least fixed point, it begins with the smallest
potential value (the set of target configurations $C$ in the case of
reachability properties, and the empty set in the case of \buchi properties).
It then adds configurations to this set (by adding new transitions) that also
necessarily appear in the least fixed point.  This process is repeated until
nothing more needs to be added -- at which point the least fixed point has been
calculated.

To compute a greatest fixed point $Z_0$ we follow the dual strategy.  We begin
with the largest possible value, which is the set of all configurations, which
we will represent by states $\pinit{p}{0}$ with all possible outgoing
transitions.  Next, the least fixed point $Z_1$ is calculated given the initial
approximation of $Z_0$.  Once the value of $Z_1$ is known, it becomes our new
approximation of $Z_0$.  Notice that this approximation is necessarily smaller
than the initial attempt (both in terms of configurations accepted and
transitions present).  We then recalculate the least fixed point for $Z_1$ with
the new smaller value of $Z_0$.  In this way, starting from the largest possible
value for $Z_0$ we successively shrink its value until a fixed point is found.
This fixed point will be the greatest fixed point.

\paragraph{Projection}

When computing the greatest fixed point for $Z_0$ we repeatedly compute a least
fixed point for $Z_1$.  Each fixed point for $Z_1$ becomes the new approximation
of $Z_0$.  Hence, during our algorithm we need a method of assigning the value
of $Z_1$ to $Z_0$.  We call this manipulation of transitions \emph{projection}.

Suppose the only outgoing transition from $\pinit{p}{1}$ is
\[
    \pinit{p}{1} \xrightarrow{A} \set{\pinit{q}{1},\pinit{p}{0}}
\]
and we want to assign the new value of $\pinit{p}{0}$.  To do this we simply
remove all transitions from $\pinit{p}{0}$ (the old value) and introduce the
transition
\[
    \pinit{p}{0} \xrightarrow{A} \set{\pinit{q}{0},\pinit{p}{0}} \ .
\]
There are several things to notice about this new transition.  The first is that
it emanates from $\pinit{p}{0}$ rather than $\pinit{p}{1}$.  Next, we have
changed the target state $\pinit{q}{1}$ to $\pinit{q}{0}$.  This is because we
are renaming all the states annotated with $1$ to be annotated with $0$.
Finally, notice that we have not changed the target state $\pinit{p}{0}$.

By leaving $\pinit{p}{0}$ we are no longer simply transferring the value of
$Z_1$ to $Z_0$ since we are changing the outgoing transitions from
$\pinit{p}{0}$.  It is provable that this change in value is benign with respect
to the fixed point of $Z_0$: since $\pinit{p}{0}$ should accept all
configurations $(p, w)$ in the fixed point for $Z_0$, the fact that any run that
reaches $\pinit{p}{0}$ may accept additional configurations coming from the new
value of $\pinit{p}{0}$ rather than the old simply means that we are
accelerating the computation of the fixed point.

For example, suppose we had a pushdown \buchi game with $p \in F \cap Q_E$ and
an automaton with the transitions
\[
    \pinit{p}{1} \xrightarrow{A} \set{\pinit{p}{0}} \text{ and } \pinit{p}{1}
    \xrightarrow{\bot} \set{\qend} \text{ and } \pinit{p}{0}
    \xrightarrow{\bot} \set{\qend}
\]
and the pushdown game contains (amongst others) the rule $(p, A) \rightarrow
(p, \varepsilon)$.  In particular we accept the configuration $(p, A\bot)$
from $\pinit{p}{1}$, and we do so because we can pop the $A$ to reach $(p,
\bot)$ (from which we suppose \eloise can win the game).  After projection, we
will have the transitions
\[
    \pinit{p}{0} \xrightarrow{A} \set{\pinit{p}{0}} \text{ and } \pinit{p}{0} \xrightarrow{\bot} \set{\qend} \ .
\]
Notice we now have a loop from $\pinit{p}{0}$ enabling any configuration of the
form $(p, A^\ast\bot)$ to be accepted from $\pinit{p}{0}$.  Thus we have
increased the valuation during projection.  However, this is benign because, by
repeated applications of $(p, A) \rightarrow (p, \varepsilon)$ \eloise can
reach $(p, \bot)$ and win the game.  Thus, the projection has collapsed an
unbounded sequence of moves into a single transition.

To calculate the fixed point for $Z_1$ we begin with the empty set as an initial
approximation.  Then we compute the new approximation for $Z_1$.  While
computing this approximation we will use states of the form $\pinit{p}{2}$ to
store the new value.  Thus, to assign the new approximation to $Z_1$ we simply
perform projection from the states $\pinit{p}{2}$ to $\pinit{p}{1}$ in the same
way that we projected when assigning $Z_1$ to $Z_0$.

We thus define a projection function on states
\[
    \proj{\gcoli}{\gcolj}(s) = \begin{cases}
        s & s = \qall \lor s = \qend \\
        s & s = \pinit{p}{\gcolk} \land \gcolk \neq \gcoli \\
        \pinit{p}{\gcolj} & s = \pinit{p}{\gcoli}
    \end{cases}
\]
which generalises naturally to a function on sets of states
$\proj{\gcoli}{\gcolj}(S) = \setcomp{\proj{\gcoli}{\gcolj}(s)}{s \in S}$.

\paragraph{Algorithm}

Fix a pushdown \buchi game $\pda = (Q, \Gamma, \bot, \Delta, F)$.  We begin our
presentation of the algorithm by presenting a simple function for performing the
projections described above.  The function \Call{\fpproj}{$\altA$, $\gcoli$,
$\gcolj$} projects the value of the states $\pinit{p}{\gcoli}$ to
$\pinit{p}{\gcolj}$ and deletes all the states $\pinit{p}{\gcoli}$.
\begin{algorithmic}
    \Function{\fpproj}{$\altA$, $\gcoli$, $\gcolj$}
         \State $(\astates, \Gamma, \adelta, \afinals) \gets \altA$

         \State $\astates' \gets \astates \setminus \setcomp{\pinit{p}{\gcoli}}{p \in
         Q}$

         \State $\adelta' \gets \begin{array}{l}
             \setcomp{s \xrightarrow{A} S \in \adelta}{\forall p \in Q . s \neq \pinit{p}{\gcoli}
             \land s \neq \pinit{p}{\gcolj}} \cup \\

             \setcomp{\pinit{p}{\gcolj} \xrightarrow{A} \proj{\gcoli}{\gcolj}(S)}{\pinit{p}{\gcoli}
             \xrightarrow{A} S \in \adelta}
         \end{array}$

         \Return {$(\astates', \Gamma, \adelta', \afinals)$}
    \EndFunction
\end{algorithmic}

The main algorithm contains two nested fixed point computations: the outer for
$Z_0$ and the inner for $Z_1$.  The initial automaton $\psatinitA$ contains only
the states $\qall$ and $\qend$ with transitions as described above.  That is
$\psatinitA = (\set{\qall, \qend}, \Gamma, \adelta, \set{\qend})$ with
\[
    \adelta = \setcomp{\qall \xrightarrow{A} \set{\qall}}{A \in \Gamma \land A
    \neq \bot} \cup \set{\qall \xrightarrow{\bot} \set{\qend}} \ .
\]
The algorithm is then a call to the function \Call{\fpstepb{0}}{$\psatinitA$}
defined below.  We define two functions for computing the fixed points for $Z_0$
and $Z_1$.  Both of these functions are similar to each other: they begin by
setting up an automaton representing the initial approximation of the fixed
point, either by adding no transitions (the empty set) or all transitions (the
largest set).  They then enter a loop of computing the next approximation and
then using projection to transfer (and accelerate) the new value to the states
$\pinit{p}{0}$ or $\pinit{p}{1}$ as appropriate.  The function
\Call{\fpstepb{0}}{$\altA$} computes the fixed point for $Z_0$ and uses
\Call{\fpstepb{1}}{$\altA$} to compute the next approximation, while
\Call{\fpstepb{1}}{$\altA$} computes the fixed point for $Z_1$ and uses a
function \Call{\fppre}{$\altA$} to compute the next approximation.  These two
functions are thus defined
\begin{algorithmic}
    \Function{\fpstepb{0}}{$\altA$}
        \State $(\astates, \Gamma, \adelta, \afinals) \gets \altA$

        \State $\astates' \gets \astates \cup \setcomp{\pinit{p}{0}}{p \in Q}$

        \State $\adelta' \gets \setcomp{\pinit{p}{0} \xrightarrow{A} S}{p \in Q
        \land A \in
        \Gamma \land A \neq \bot \land S \subseteq \astates' \setminus
        \set{\qend}} \cup \setcomp{\pinit{p}{0} \xrightarrow{\bot}
        \set{\qend}}{p \in Q}$

        \State $\altB \gets (\astates', \Gamma, \adelta', \afinals)$

        \Repeat
            \State $\altB \gets$ \Call{\fpstepb{1}}{$\altB$}

            \State $\altB \gets$ \Call{\fpproj}{$\altB$, 1, 0}
        \Until {$\altB$ unchanged}

        \Return $\altB$
    \EndFunction
\end{algorithmic}
and
\begin{algorithmic}
    \Function{\fpstepb{1}}{$\altA$}
        \State $(\astates, \Gamma, \adelta, \afinals) \gets \altA$
        \State $\astates' \gets \astates \cup \setcomp{\pinit{p}{1}}{p \in Q}$

        \State $\altB \gets (\astates', \Gamma, \adelta, \afinals)$

        \Repeat
            \State $\altB \gets$ \Call{\fppre}{$\altB$}

            \State $\altB \gets$ \Call{\fpproj}{$\altB'$, 2, 1}
        \Until {$\altB$ unchanged}

        \Return $\altB$
    \EndFunction \ .
\end{algorithmic}
The inner fixed point computation uses a function \Call{\fppre}{$\altA$} to
compute the step of the calculation corresponding to
\[
    \brac{F \cap \prestep{\pda}{Z_0}} \cup
    \brac{\negation{F} \cap \prestep{\pda}{Z_1}} \ .
\]
This function adds transitions in the same way as the loop of saturation
algorithm for reachability games, except it is sensitive to the two different
fixed point variables.  For convenience, we define the function $\Omega$ such
that
\[
    \Omega(p) = \begin{cases}
                    0 & p \in F \\
                    1 & p \notin F \ .
                \end{cases}
\]
We can then define
\begin{algorithmic}
    \Function{\fppre}{$\altA$}
        \State $(\astates, \Gamma, \adelta, \afinals) \gets \altA$

        \State $\astates' \gets \astates \cup \setcomp{\pinit{p}{2}}{p \in Q}$

        \State $\adelta' \gets \begin{array}{l}
            \setcomp{\pinit{p}{2} \xrightarrow{A} S}{
                p \in Q_E \land
                \exists (p, a) \rightarrow (q, w) \in \Delta .
                    \pinit{q}{{\Omega(p)}} \xrightarrow{w} S
            } \cup \\
            \setcomp{\pinit{p}{2} \xrightarrow{A} S}{
                p \in Q_A \land
                \forall (p, a) \rightarrow (q, w) \in \Delta .
                    \exists \pinit{q}{{\Omega(p)}} \xrightarrow{w} S' .
                        S' \subseteq S
            }
        \end{array}$

        \Return $(\astates', \Gamma, \adelta', \afinals)$
    \EndFunction \ .
\end{algorithmic}

The automaton $\altB$ that is the result of \Call{\fpstepb{0}}{$\psatinitA$}
will be such that $(p, w) \in \winningregion$ iff $w \in
\slang{\pinit{p}{0}}{\altB}$.  Since there are at most an exponential number of
transitions in the automaton each fixed point may iterate at most an exponential
number of times.  This gives us an overall exponential run time for the
algorithm.
\begin{theorem}
    The winning region of a pushdown \buchi game is regular and computable in
    exponential time.
\end{theorem}

Note that for the one player case (\emph{i.e.} all states belong to \eloise), the computation can be done in polynomial time \cite{BEM97,FWW97}.

\subsubsection{Winning strategies}

Cachat also showed that, like in reachability games, it is possible to construct
a linear space positional strategy and a constant time (though not positional)
pushdown strategy for \eloise.  However, in his PhD. thesis~\cite{Cachat03}
Cachat observes that adopting his techniques for computing strategies for
\abelard is not clear.  However, it is known that, even for the full case of
parity games, a pushdown strategy exists using different
techniques~\cite{W01,Serre04}.

\subsection{Pushdown parity games}

Parity games allow more complex liveness properties to be checked.  To
define a parity game, each configuration is assigned a ``colour'' from a set of
colours represented by natural numbers.  The winner of the game depends on the
smallest colour appearing infinitely often in the run: if it is even then
\eloise wins the game, else \abelard wins.

More formally, given a sequence of configurations $\rho = (q_0, w_0), (q_1,
w_1), \ldots$ let $\infocc(\rho)$ be the set of control states appearing
infinitely often in $\rho$.  That is
\[
    \infocc(\rho) = \setcomp{q}{\forall i \exists j > i . q_j = q} \ .
\]

Given a set of control states $Q$ and maximum colour $\gmaxcol$, let $\Omega : Q
\rightarrow \set{0,\ldots,\gmaxcol}$ be a colouring function assigning colours to each
control state.  We can generalise $\Omega$ to sets of control states $P$ by
taking the image of $P$.  That is, $\Omega(P) = \setcomp{\gcoli}{\exists p \in
P .  \Omega(p) = \gcoli}$.

A pushdown parity game is a tuple $(Q, \Gamma, \bot, \Delta, \Omega)$ where
$\Omega : Q \rightarrow \set{0,\ldots,\gmaxcol}$ is a colouring function
assigning to each control state a colour from the set
$\set{0,\ldots,\gmaxcol}$.  Moreover, the tuple defines a pushdown game $(Q,
\Gamma, \bot, \Delta, \wincond)$ where
\[
    \wincond = \setcomp{\rho}{\minf(\Omega(\infocc(\rho))) \text{ is even}} \ .
\]

Thus, a \buchi game is a special case of a parity game, where the set of colours
is $\set{0, 1}$ and
\[
    \Omega(p) = \begin{cases}
                    0 & p \in F \\
                    1 & p \notin F \ .
                \end{cases}
\]

\subsubsection{Characterising the winning region}

The characterisation of \eloise's winning region in terms of fixed points is a
natural extension of the \buchi version.  That is, assuming $\gmaxcol$ to be odd
and writing $C_\gcoli$ to denote $\setcomp{(p, w)}{\Omega(p) = \gcoli}$, we need
\[
    \nu Z_0 . \mu Z_1 . \cdots . \nu Z_{\gmaxcol-1} . \mu Z_\gmaxcol .  \bigcup\limits_{0 \leq
    \gcoli \leq \gmaxcol} \brac{C_\gcoli \cap \prestep{\pda}{Z_\gcoli}} \ .
\]
This formula can be understood as a generalisation of the \buchi formula, where
$F = C_0$ and $\negation{F} = C_1$.  When the colour of a configuration is odd,
then it is bound by a least fixed point.  Hence, it must eventually exit this
fixed point by visiting a configuration with a smaller colour (just like a
configuration in $\negation{F}$ had to visit a configuration in $F$).  When the
colour is even, then it is bound by a greatest fixed point -- hence a play can
stay within this fixed point, never visiting a smaller colour, and satisfy the
winning condition for \eloise.

\subsubsection{Computing the winning region}

Fix a pushdown parity game $\pda = (Q, \Gamma, \bot, \Delta, \Omega)$.
Computing the winning region in a pushdown parity game is a direct extension of
the algorithm presented for \buchi games.  Since a \buchi game is simply a
pushdown parity game with two colours, we generalise the nesting of the fixed
point calls to an arbitrary number of colours.  To this end we introduce a
function \Call{\fpdispatch}{$\altA$, $\gcoli$} that manages the level of
nesting, and performs a fixed point or a pre-step analysis as appropriate.
\begin{algorithmic}
    \Function{\fpdispatch}{$\altA$, $\gcoli$}
        \If {$\gcoli = \gmaxcol + 1$}
            \State \Return \Call{\fppre}{$\altA$}
        \Else
            \State \Return \Call{\fpstepp}{$\altA$, $\gcoli$}
        \EndIf
    \EndFunction
\end{algorithmic}
Using this function we can define a generic fixed point function based on the
\buchi functions.  This function performs the nested calculations and the
projection as before.  The initial transitions from the new states introduced by
the function depend on the parity of $\gcoli$: when computing an even (greatest)
fixed point, we add all transitions, and when computing an odd (least) fixed
point, we add no transitions.
\begin{algorithmic}
    \Function{\fpstepp}{$\altA$, $\gcoli$}
        \State $(\astates, \Gamma, \adelta, \afinals) \gets \altA$

        \State $\astates' \gets \setcomp{\pinit{p}{\gcoli}}{p \in Q}$

        \If {$\gcoli$ is even}
            \State $\adelta' \gets \setcomp{\pinit{p}{\gcoli} \xrightarrow{A} S}{p \in
            Q \land A \in \Gamma \land A \neq \bot \land S \subseteq \astates'
            \setminus \set{\qend}} \cup \setcomp{\pinit{p}{\gcoli}
            \xrightarrow{\bot} \set{\qend}}{p \in Q}$
        \Else
            \State $\adelta' \gets \emptyset$
        \EndIf

        \State $\altB \gets (\astates \cup \astates', \Gamma, \adelta \cup \adelta', \afinals)$

        \Repeat
            \State $\altB \gets$ \Call{\fpdispatch}{$\altB$, $\gcoli + 1$}

            \State $\altB \gets$ \Call{\fpproj}{$\altB$, $\gcoli + 1$, $\gcoli$}
        \Until {$\altB$ unchanged}

        \Return $\altB$
    \EndFunction
\end{algorithmic}
Finally, we redefine the \Call{\fppre}{$\altA$} function to add transitions to
the correct initial states.  Note, we were already using $\Omega$ to distinguish
between different fixed point variables, hence this function is almost identical to the \buchi case.
\begin{algorithmic}
    \Function{\fppre}{$\altA$}
        \State $(\astates, \Gamma, \adelta, \afinals) \gets \altA$

        \State $\astates' \gets \astates \cup \setcomp{\pinit{p}{{\gmaxcol+1}}}{p \in Q}$

        \State $\adelta' \gets \begin{array}{l}
            \setcomp{\pinit{p}{{\gmaxcol+1}} \xrightarrow{A} S}{
                p \in Q_E \land
                \exists (p, a) \rightarrow (q, w) \in \Delta .
                    \pinit{q}{{\Omega(p)}} \xrightarrow{w} S
            } \cup \\
            \setcomp{\pinit{p}{{\gmaxcol+1}} \xrightarrow{A} S}{
                p \in Q_A \land
                \forall (p, a) \rightarrow (q, w) \in \Delta .
                    \exists \pinit{q}{{\Omega(p)}} \xrightarrow{w} S' .
                        S' \subseteq S
            }
        \end{array}$

        \Return $(\astates', \Gamma, \adelta', \afinals)$
    \EndFunction
\end{algorithmic}
Thus, to compute the winning region of a pushdown parity game, we make the call
\Call{\fpdispatch}{$\psatinitA$, 0} where $\psatinitA$ is the initial automaton
with only the states $\qall$ and $\qend$ as defined in the \buchi case.

The automaton $\altB$ that is the result of \Call{\fpdispatch}{$\psatinitA$, 0}
will be such that $(p, w) \in \winningregion$ iff $w \in
\slang{\pinit{p}{0}}{\altB}$.  Since there are at most an exponential number of
transitions in the automaton each fixed point may iterate at most an exponential
number of times.  This gives us an overall exponential run time for the
algorithm.
\begin{theorem}
    The winning region of a pushdown parity game is regular and computable in
    exponential time.
\end{theorem}

\subsubsection{Winning strategies}

Unfortunately, it is currently unknown how to compute the winning strategies for
\eloise and \abelard using the saturation technique for pushdown parity games.
However, using a different approach, both Walukiewicz~\cite{W01} and
Serre~\cite{Serre04} have shown that a pushdown strategy exists for both
players.


%% file: implementations.tex
\section{Implementations and Applications of Saturation Methods}
\label{sec:implementations}

In this article, we have presented the saturation method from a theoretical
standpoint.  The method, however, is an algorithmic approach that is well suited
to implementation, and several tools have been constructed using saturation as
its core technique.

\subsection{Single Player Implementations}

Perhaps the most famous of these tools is Moped~\cite{ES01,S02} and its
incarnation as a model checker for Java, JMoped~\cite{SSE05,SBSE07}.  In taking
the algorithm from a theoretical tool to a practical one, a number of new
concerns had to be taken into account.

The rules of a pushdown system roughly correspond to the statements in a
program.  In a program with thousands of lines, a fixed point iteration that
checks, during each iteration, whether each rule leads to new transitions in
the automaton would be woefully inefficient.  In constructing Moped,
Esparza\etal~\cite{EHRS00} showed how this naive outer loop can be reorganised
such that, at each iteration, only the relevant rules of the system were
considered, leading to a significant improvement in performance.

A second consideration of applications to the analysis of program models is the
handling of data values.  Boolean programs are essentially pushdown systems
where each control state and stack character contains a valuation of a set of
global and local boolean variables respectively.  These boolean programs are
the natural output of predicate abstraction tools such as SATABS~\cite{CKSY05}
as well as the target compilation language of JMoped.

Since there are only finitely many valuations of sets of boolean variables, they
can directly be encoded as control states or characters and standard pushdown
analysis techniques can be employed.  However, since they are also exponential
in number, such an approach is inherently inefficient.  Hence, Esparza\etal
introduced \emph{symbolic pushdown systems}~\cite{ES01} which make boolean
valuations first class objects.  The saturation technique was extended by adding
BDDs representing variable valuations to the edges of the $\pda$-automata,
leading to an implementation capable of analysing symbolic pushdown systems
derived from real-world programs.

Around this time it was observed by Reps that the BDDs could be replaced by any
abstract domain of values that was sufficiently well behaved, and many static
analyses could be derived.  This led to the introduction of \emph{weighted
pushdown systems}~\cite{RSJM05} (and, indeed, \emph{extended} weighted pushdown
systems amongst other improvements~\cite{LRB05,LR06}), of which symbolic
pushdown systems and their BDD representation were an instance.  The developers
of Moped created the \emph{weighted pushdown system library}~\cite{wpdsLib} as a
component of Moped, and Reps\etal developed WALi~\cite{WALi} implementing these
new algorithms.

\subsection{Two-Player Implementations}

Perhaps the most straight-forward optimisation to make to the saturation
technique as presented for two-player games is via the observation that a
transition
\[
    s \xrightarrow{A} S
\]
is effectively redundant if there exists another transition
\[
    s \xrightarrow{A} S'
\]
with $S' \subseteq S$.  This is because an accepting run from $S$ contains
within it an accepting run from $S'$, and thus the former transition can be
removed.

When considering reachability games, it is also possible to improve the naive
fixed point iteration, as in the single-player case, to avoid checking against
all pushdown rules during each step of the implementation.  Such an optimisation
was introduced by Suwimonteerabuth\etal and implemented with applications to
certificate chain analysis~\cite{SSE06}.

This work has recently been built upon by Song who has developed various tools
based upon reductions to \buchi games and tools for their analysis.  Primarily
this work has focussed on a specification language that is an extension of CTL
and its translation into symbolic pushdown \buchi games~\cite{ST11,ST12}
resulting in the tool PuMoC~\cite{ST12b}.  The main application of this work has
been in the detection of malware.  More recently still, this work has been
developed for LTL-like properties to deal with situations where the CTL approach
was insufficient~\cite{ST13}, culminating in the PoMMaDe tool~\cite{ST13b}.

However, the combination of BDD representations and alternating automata is not
an easy one, since BDDs lack the necessary alternation for a direct embedding.
Hence, Song's algorithm pays an extra exponential in its worst-case complexity
(doubly exponential rather than exponential), although the practical runtime is
improved.  The optimal inclusion of symbolic representations into the analysis
of two-player games remains an open problem.

The saturation technique for the full case of parity games has been implemented
in the PDSolver tool~\cite{HO10} and applied to dataflow analysis problems for
Java programs.  Due to the interactions between the several layers of fixed
points, it is not clear how to adapt Esparza\etal and Suwimonteerabuth\etal's
efficient algorithms to this case, nor how to include symbolic representations.
These remain limitations of the tool, and interesting avenues for future work.


%% file: others.tex
\section{Extensions of the Saturation Method}
\label{sec:others}

In this article we have looked at the different saturation methods for pushdown
systems.  Across several articles, the technique has proved to be applicable to
various extensions to the basic model.  We briefly list some of these results
here.

\paragraph{Concurrency}

The reachability problem for pushdown systems with two or more stacks is well
known to be undecidable.  Since multiple stacks are needed to model multi-thread
recursive programs, a number of underapproximation techniques have been studied
for which the reachability problem is decidable.  One such technique is
\emph{bounded context switching}~\cite{Q08} where the number of interactions
between the threads is limited to an \textit{a priori} fixed number $k$.  While
this cannot prove the absence of errors, it is effective at finding bugs in
programs, since, empirically, bugs usually manifest themselves within a small
number of interactions.  This restriction can be relaxed further by allowing a
bounded number of \emph{phases}~\cite{lTMP07} (where all threads run
concurrently, but during each phase only one thread is allowed to pop from its
stack), or a bounded scope~\cite{lTN11} (where, threads are scheduled in a
round-robin fashion, and characters may only be removed from the stack if they
were pushed at most a fixed number of rounds earlier).

The saturation technique has proved useful for each of these restrictions.  In
particular, Moped has been extended to provide context bounded analysis of
multi-stack pushdown systems~\cite{SES08} by Suwimonteerabuth\etal and
saturation was used by Seth to provide a regular solution to the global
reachability problem for phase bounded pushdown systems~\cite{S10}.  The
original proof that the reachability problem for scope bounded pushdown systems
is decidable was itself an extension of the saturation technique~\cite{lTN11}.

An alternative restriction that permits a decidable reachability and LTL model
checking problem is that of \emph{ordered multi-pushdown systems} where only the
leftmost non-empty stack is able to remove characters.  Atig provides two
extensions of the saturation technique in this direction~\cite{A12}.  First,
instead of each pushdown rule adding a fixed sequence of characters to the
stack, he allows rules to contain languages of sequences that may be pushed.  If
it is decidable whether the language of a rule intersected with a regular
language is empty, then an augmented saturation technique leads to an effective
analysis algorithm.  In particular, the model checking problem for ordered
pushdown systems can be solved with this formalism.

Finally, Song generalises his LTL model checking algorithms to the case of
pushdown systems with dynamic thread creation~\cite{ST13c}, again using a
saturation technique at its core.

\paragraph{Ground Tree Rewrite Systems and Resources}

Ground tree rewrite systems can be thought of as pushdown systems with a single
control state and a more complex stack structure.  That is, the stack is a tree
rather than a word.  Rewrite rules in this system replace complete subtrees.
For example a push rule $(p, A) \rightarrow (p, BC)$ can be considered to be
replacing the subtree consisting in the leaf node $A$ with the subtree $B(C)$
(i.e. a $B$-node with a $C$-leaf as a child).  In 1987, Dauchet\etal used
saturation to show that the confluence problem for these systems is
decidable~\cite{DTHL87}.  More recently, Lang and L\"oding adapted this method
to analyse prefix replacement systems with resource usage~\cite{LL13}.

\paragraph{Higher-Order and Collapsible Pushdown Systems}

Pushdown systems provide a natural model for first-order recursive programs.
When considering higher-order programs, we can use \emph{higher-order pushdown
systems}~\cite{M76} whose stacks have a nested ``stack-of-stacks'' structure.
These systems correspond to \emph{higher-order recursion schemes} satisfying a
\emph{safety} constraint~\cite{KNU02}.  Recently, these systems were generalised
to \emph{collapsible pushdown systems} (via \emph{panic
automata}~\cite{KNUW05}), providing an automata model without the need for the
safety constraint~\cite{HMOS08}.

The saturation technique was first applied to the analysis of higher-order
systems by Bouajjani and Meyer~\cite{BM04} who considered higher-order pushdown
systems with a single control state.  This algorithm was generalised by Hague
and Ong to permit an arbitrary number of control states~\cite{HO08}.  An
alternative construction in the case of second order higher-order pushdown
systems was provided by Seth~\cite{S08}.

More recently this approach was developed by Broadbent\etal to obtain a
saturation algorithm for the full case of collapsible pushdown
systems~\cite{BCHS12}, leading to the analysis tool \cshore~\cite{BCHS13}.  This
algorithm was applied directly to the analysis of recursion schemes (without the
intermediate automata model) by Broadbent and Kobayashi, resulting in the
\horsat tool~\cite{BK13}.

Finally, the case of concurrent higher-order systems has been briefly
considered.  Seth used saturation to show that parity games over phase-bounded
higher-order pushdown systems (without collapse) are effectively
solvable~\cite{S09}.  Recently, Hague showed that the saturation approaches for
first-order phase-bounded, ordered and scope-bounded pushdown systems can be
adapted to solve the analogous reachability problems for collapsible pushdown
systems~\cite{H13}.
